\documentclass[%
 reprint,
superscriptaddress,
 amsmath,amssymb,
 aps,
 prl
]{revtex4-2}

\usepackage{graphicx}
\usepackage{dcolumn}
\usepackage{bm}
\usepackage{soul}

\usepackage[colorlinks]{hyperref}

\usepackage[english]{babel}
\usepackage{color}
\usepackage{mathtools}
\usepackage{cleveref}

\usepackage{verbatim}
\usepackage{breqn}

\makeatletter
\newcommand{\customlabel}[2]{%
   \protected@write \@auxout {}{\string \newlabel {#1}{{#2}{\thepage}{#2}{#1}{}} }%
   \hypertarget{#1}{}
}
\makeatother

\begin{document}

\title{Self-Pulsing in driven-dissipative photonic Bose-Hubbard dimers}

\author{Jes\'us Yelo-Sarri\'on}

\author{Pedro Parra-Rivas}

\author{Nicolas Englebert}

\author{Carlos Mas Arab\'i}

\author{Fran\c{c}ois Leo}

\author{Simon-Pierre Gorza}

\affiliation{OPERA-Photonique$,$ Université libre de Bruxelles$,$ 50 Avenue F. D. Roosevelt$,$ CP 194/5 B-1050 Bruxelles$,$ Belgium}

\begin{abstract}
We experimentally investigate the nonlinear dynamics of two coupled fiber ring resonators, coherently driven by a single laser beam.  
We comprehensively explore the optical switching arising when scanning the detuning of the undriven cavity, and show how the driven cavity detuning dramatically changes the resulting hysteresis cycle. By driving the photonic dimer out-of-equilibrium, we observe the occurrence of stable self-switching oscillations near avoided resonance crossings.
All results agree well with the driven-dissipative Bose-Hubbard dimer model in the weakly coupled regime.

\end{abstract}

\maketitle

The spontaneous emergence of sustained periodic oscillations is a fascinating and ubiquitous phenomenon arising in nonlinear systems. It is associated with the breaking of translational symmetry in time, and is encountered in various fields as diverse as chemistry, biology, mechanical engineering, or astrophysics to cite only a few\,\cite{Nicolis1995, Jenkins2013}. 
Since the seminal works of Lotka\,\cite{Lotka1920} and Volterra\,\cite{Volterra1926} a century ago in the context of chemistry and population dynamics in biology, respectively, it is now well known that such undamped oscillations may arise 
in nonlinear systems with coupled variables under continuous driving.
An important example is the generation of infinite trains of pulses in the FitzHugh–Nagumo model of nerve membranes\,\cite{FitzHugh1961}. 
In optics, we can cite for instance self-pulsing in second-harmonic generation\,\cite{Drummond1980}, in lasers between coupled longitudinal modes\,\cite{Paoli1969} or with continuous injected signal\,\cite{Lugiato1983} or, more recently, between counterpropagating beams in a single Kerr ring resonator\,\cite{Woodley2020}. 
Rhythmogenesis refers to the emergence of oscillations from the coupling between two or more sub-systems that show only steady states when uncoupled\,\cite{Epstein1996}. 
In this context, the driven-dissipative Bose-Hubbard (DDBH) model plays an essential role in physics, as it provides a canonical description of the dynamics between strongly interacting bosons for open quantum systems\,\cite{Bruder2005}.  
In its simplest realization, only two macroscopic phase coherent wave functions are coupled to form a Bose-Hubbard dimer, also referred as a bosonic Josephson junction. 
These junctions have been initially investigated with superconductors separated by a thin insulator\,\cite{Likharev1979} and with coupled reservoirs of super-fluid helium\,\cite{Pereverzev1997}. Later, it was realized that they can be implemented with weakly coupled Bose-Einstein condensates in a macroscopic double-well potential\,\cite{Albiez2005} and in photonic systems with coupled semiconductor microcavities hosting polariton excitation\,\cite{Abbarchi2013}. However, beside Josephson effects, owing to their intrinsic nonlinearity, other striking phenomena such as anharmonic oscillations or macroscopic quantum self-trapping emerge in these latter systems
\,\cite{Albiez2005, Abbarchi2013, Smerzi1997, Sarchi2008}.        
Interestingly, under continuous excitation, it was theoretically predicted that sustained oscillations may take place in DDBH systems. This was shown for microcavity polaritons\,\cite{Sarchi2008}, but also for nonlinear optical cavities with two \cite{Eksioglu2015, Grigoriev2011} or more\,\cite{Maes2009, Petracek2014, Armaroli2015} coupled cavities. 
It is however only very recently that evidence of such self-pulsing was reported with polaritons, through its indirect spectral signature\,\cite{Zambon2020}. 
Moreover, coupled ring resonators host rich nonlinear dynamics and have recently attracted a lot of attention for frequency-comb generation in micro-resonators (see e.g.\,\cite{Xue2015, Miller2015, Fujii2018, Xue2019, Tikan2020, Helgason21}). 

In this Letter, we report, for the first time to our knowledge, a comprehensive experimental investigation of the dynamical regimes of driven-dissipative photonic dimers and the observation of the spontaneous emergence of sustained oscillations between light beams propagating in two linearly coupled ring resonators. We consider passive fiber cavities corresponding to asymmetrically excited photonic DDBH dimers. We focus on weak coupling in relation to dissipation, which is relevant for moderate to low finesse resonators. In photonic dimers, tuning the cavity detunings is analogous to changing the single particle energy of the two quantum states in bosonic Josephon junctions\,\cite{Abbarchi2013, Rodriguez2016}. Here, contrary to integrated dimers, the two cavity detunings can be independently set or scanned without linear or nonlinear couplings between them.   

\begin{figure}
\includegraphics[width=\columnwidth]{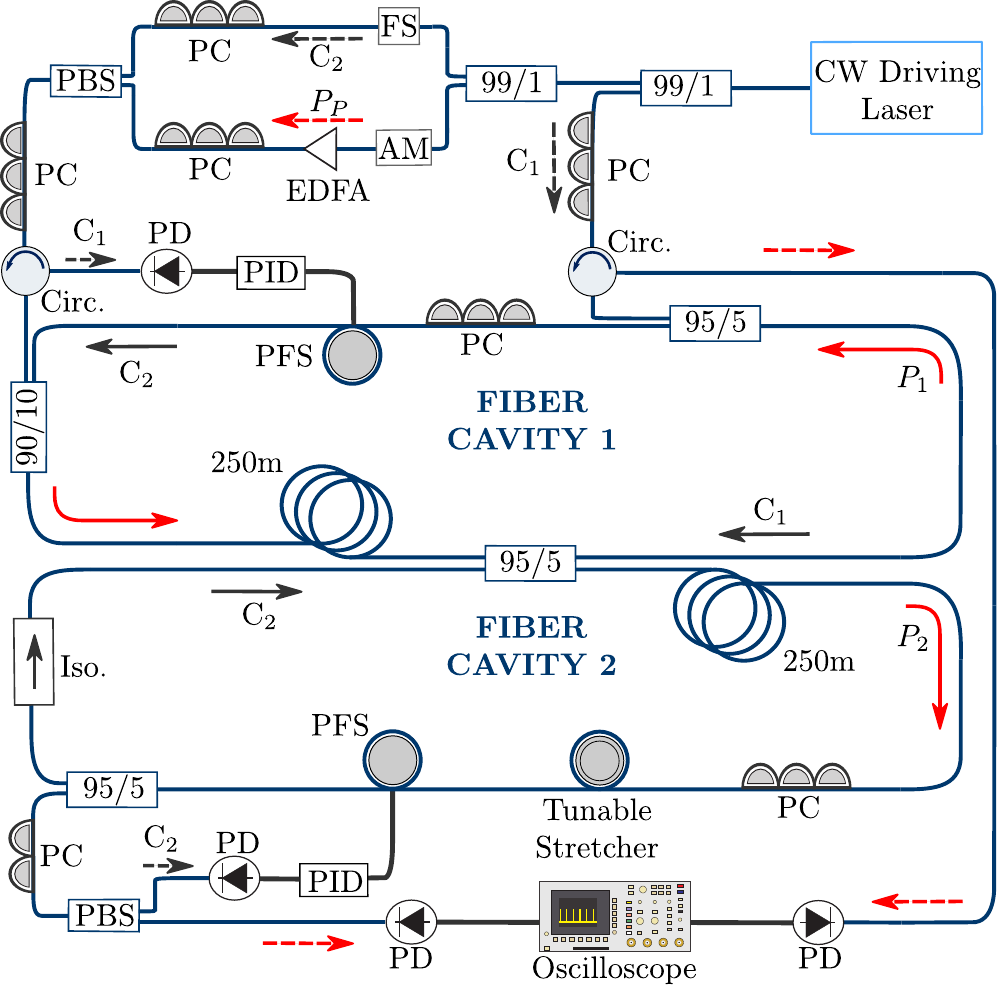}
\caption{Experimental set-up. The photonic dimer is synchronously driven by a $1550\,\mathrm{nm}$ pulsed laser beam injected in the first cavity. The detunings in each cavity are independently stabilized. AM, amplitude modulator; EDFA, erbium-doped fiber amplifier; FS, frequency shifter; PBS, fiber polarization beam splitter; PC, polarization controller; PFS, piezo-electric fiber stretcher; PD, photodiode; PID, proportional-integral-derivative controller; Iso., optical isolator; Circ., optical circulator, $\mathrm{C}_1$ (resp. $\mathrm{C}_2)$ control signal to lock $\delta_1$ (resp. $\delta_2$). The tunable stretcher is used to match the two cavity round-trip times. 
}\label{Fig1_ExpSetup}
\end{figure}

\begin{figure}
\customlabel{fig:2a}{2(a)}
\customlabel{fig:2b}{2(b)}
\customlabel{fig:2c}{2(c)}
\customlabel{fig:2d}{2(d)}
\includegraphics[width=\columnwidth]{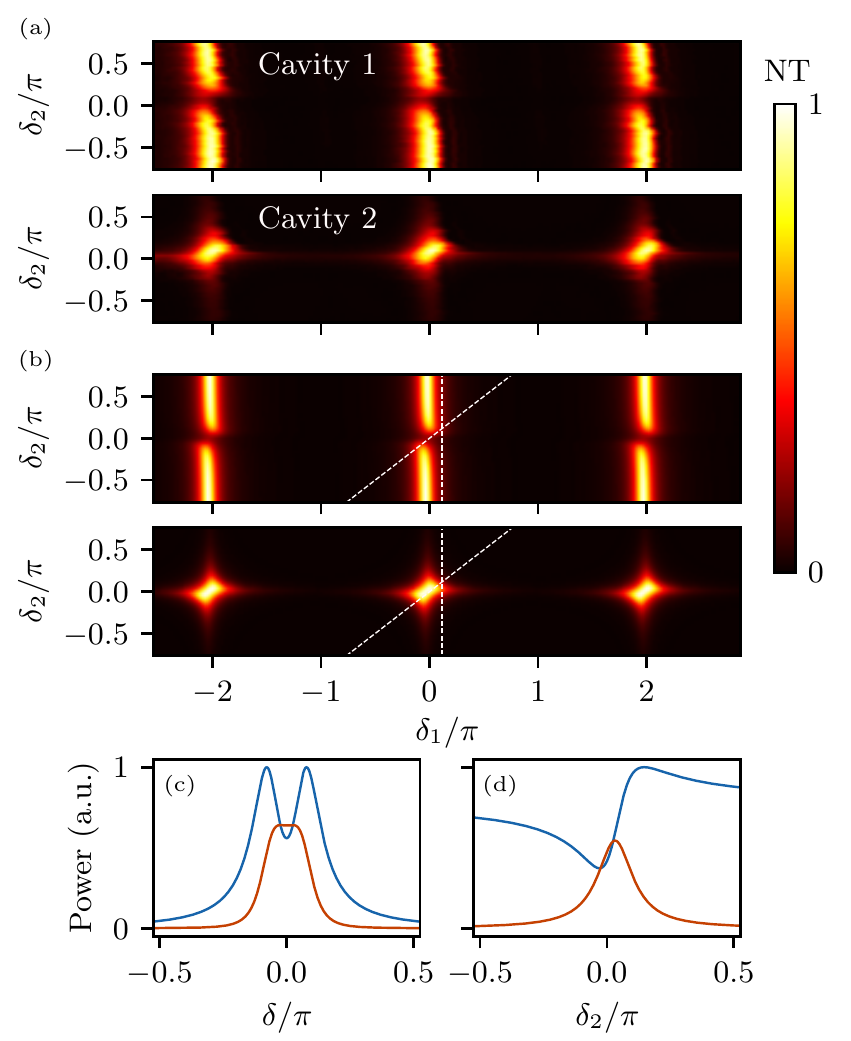}
\caption{(a,\,b) Linear resonances as a function of the two cavity detunings, in the driven (top, Cavity 1) and undriven (bottom, Cavity 2) cavities. 
(a) Experimental results for $P_{p}\approx5\,\mathrm{mW}$ and (b) theoretical resonances derived from Eq.~(\ref{MF2Ring}) but repeated every $2\pi$ for $\delta_1$.   
Normalized intracavity power in the driven (blue) and undriven (red) cavity for $\delta=\delta_1=\delta_2$ (c), and (d) for $\delta_1=0.36$ while scanning the detuning in the undriven cavity [see also dashed lines in panel (b)]. NT, normalized transmission; a.u., arbitrary units.
}
\end{figure}
The experimental setup is depicted in Fig.\,\ref{Fig1_ExpSetup}. At its heart are two passive fiber cavities coupled by a 95/5 coupler. Each resonator is composed of about $250\,\mathrm{m}$ of optical fiber ($1.14\,\mu\mathrm{s}$ round-trip time) with a net normal dispersion. The loss in each cavity is $\sim40\,\%$ (excluding the shared coupler).    
The first cavity is synchronously driven through a 90/10 coupler by flat-top $470\,\mathrm{ps}$ pulses generated from a narrow linewidth distributed feedback laser.
This prevents the build up of the Brillouin scattering radiation and lowers the intracavity average power needed to reach the self-pulsing regime. The two cavity detunigs can be independently stabilized by means of piezo-electric fiber stretchers. 
The intracavity powers are measured at the two drop ports and simultaneously recorded on an oscilloscope.
The dynamics of the slowly varying envelope of the intracavity fields $A_{1,2}$ can be described by a set of two coupled Lugiato-Lefever equations\,\cite{Lugiato1983, Xue2019}. In single ring Kerr resonators with normal dispersion, modulation instability occurs in a narrow range of parameters beyond the bistability threshold\,\cite{Haelterman1992}. In addition, different works have recently shown that in coupled ring resonators, the local dispersion, induced by mode coupling, changes the instability spectrum and allows for the generation of stable localized patterns\,\cite{Miller2015, Xue2015, Fujii2018}. 
In this work, we carefully adjust the two cavity lengths to avoid such instability by ensuring they both have the same FSR. The temporal walk-off as well as the group velocity dispersion are thus neglected in the model.
Under this simplification, the dynamics of the system is governed by the DDBH dimer model\,\cite{ Casteels2017, Giraldo2020,Jebali:20}. It describes the fields evolution with the round trips $\phi/2\pi$ in each cavity of length $L$ and reads:
\begin{dmath}\label{MF2Ring}
$$\displaystyle
2\pi\frac{\mathrm{d} A_{1,2}}{\mathrm{d}\phi}=[-\kappa -i\delta_{1,2}+i\gamma L|A_{1,2}|^2]A_ {1,2}
+i\sqrt{\theta_{12}}A_{2,1}\nonumber
+i\sqrt{\theta_{p}}F_{1,2}.$$
\end{dmath}
The detunings from the closest (single cavity) resonances are $\delta_j=m_j2\pi-\varphi_j~ (j=1,2)$, with $\varphi_j$, the round-trip linear phase shift and $m_j$ an integer number. $\kappa=0.21$ is the cavity loss coefficient and the nonlinear parameter is $\gamma=3\times 10^{-3}\,\mathrm{(Wm)}^{-1}$. Finally, $\theta_{12}=0.05$ and $\theta_{p}=0.1$ are the transmission coefficients of the middle and input couplers, respectively, and $F_1=\sqrt{P_p}$ is the driving field amplitude while $F_2=0$. The field amplitudes are normalized such that the intracavity powers (expressed in W) are given by $|A_j|^2=P_j$.  

\begin{figure}
\customlabel{fig:3a}{3(a)}
\customlabel{fig:3b}{3(b)}
\customlabel{fig:3c}{3(c)}
\customlabel{fig:3d}{3(d)}
\customlabel{fig:3e}{3(e)}

\includegraphics[width=\columnwidth]{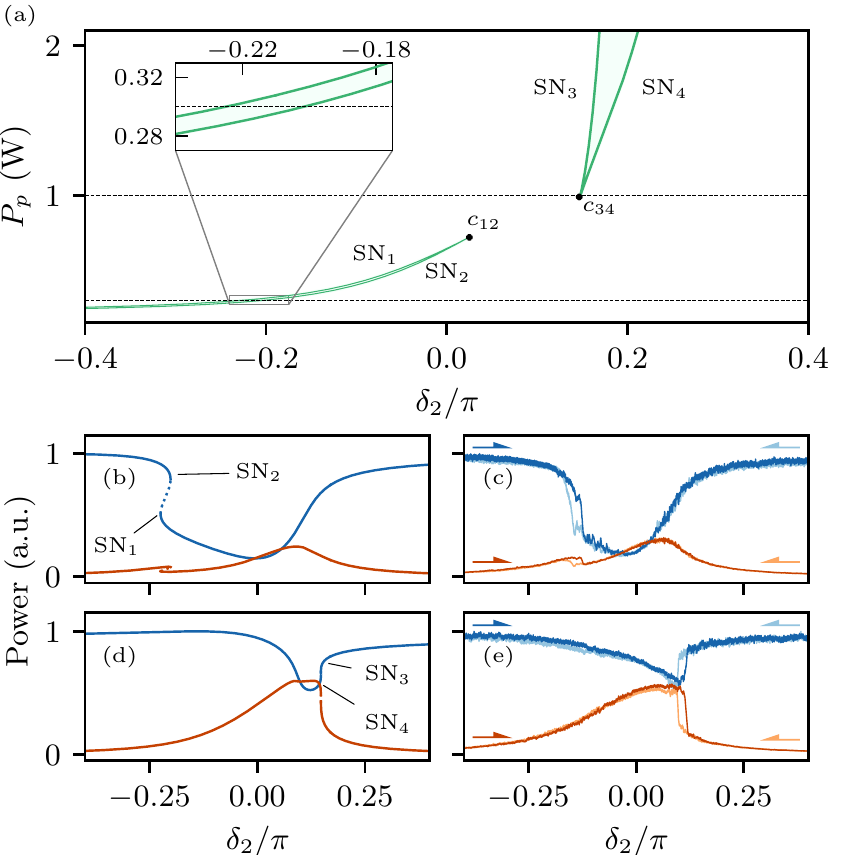}
\caption{(a) Phase diagram in the ($\delta_2,\,P_p$) parameter space for $\delta_1=0.36$. 
The saddle-node bifurcations $\mathrm{SN}_j$ are plotted as green lines and $c_{ij}$ mark cusp bifurcations. The dimer is bistable in the shaded green areas. 
(b) Bifuraction diagram showing the normalized intracavity power in the driven (blue) and undriven (red) resonators as a function of $\delta_2$, for $P_p=300\,\mathrm{mW}$ and (d) $1\,\mathrm{W}$ [see also dashed lines in panel (a)]. Solid (dashed) lines represent stable (unstable) solutions of Eq.~(\ref{MF2Ring}). (c, e) Corresponding experimental forward and backward scans. The arrows show the direction of the scans.} 
\label{fig3_delta1-036}
\end{figure}

The recorded and simulated linear resonances
are shown in Figs.~\ref{fig:2a} and \ref{fig:2b}. 
The driven cavity resonances, when scanning $\delta_1$, are reminiscent of the ones of a single ring resonator, providing that the undriven cavity is out of resonance.
When both cavities are close to resonance, the coupling splits the resonance and leads to an avoided crossing.
The resulting double peak response, usually observed when scanning the driving laser frequency\,\cite{Miller2015} [i.e. for $\delta_1=\delta_2$, see Fig.~\ref{fig:2c}], corresponds to the excitation of the bonding-like and the antibonding-like modes of the dimer\,\cite{Abbarchi2013}. However, owing to the weak coupling with respect to the loss, these two peaks are not separated here. Fig.~\ref{fig:2d} shows the intracavity powers at a fixed detuning $\delta_1=0.36$ when scanning $\delta_2$. We note that because of the resonance splitting, for $\delta_1>0$, the driving field is further from the resonance for negative than for positive $\delta_2$. It follows a non-symmetric response with $\delta_2$ that has consequences on the bifurcation diagram.        

We now investigate the nonlinear regime. 
To compute the different attractors of the system, static and dynamical, as well as their stability with respect to its main parameters, we perform a bifurcation analysis of Eq.~(\ref{MF2Ring}), following a numerical path-continuation approach using AUTO-07p\,\citep{Doedel2009}. First, we consider $\delta_1=0.36$, a value close to the detuning threshold $(\sqrt{3}\kappa)$ to observe bistability in single-ring resonators\,\cite{Coen98}. While self-pulsing is not expected at this detuning, it is an interesting value to show how the Kerr nonlinearity alters the photonic dimer resonances.  
By increasing the driving power, when scanning $\delta_2$, an optical bistablility is initially encountered in a narrow detuning range between the saddle-node bifurcation lines $\mathrm{SN}_1$ and $\mathrm{SN}_2$ in the phase diagram shown in Fig.~\ref{fig:3a}. 
The theoretical resonances at a driving power $P_p=300\,\mathrm{mW}$ are plotted in Fig.~\ref{fig:3b} and compared with the experimental transmissions reported in Fig.~\ref{fig:3c}. We observe an hysteresis cycle on the left side of the resonance in agreement with the fixed nodes of Eq.~(\ref{MF2Ring}). 
Beyond the driving power at which $\mathrm{SN}_1$ and $\mathrm{SN}_2$ meet at the cusp bifurcation $c_{12}$, the system becomes monostable again, until the emergence of a new pair of saddle-nodes $\mathrm{SN}_3$ and $\mathrm{SN}_4$ through a second cusp bifurcation $(c_{34})$. Theoretical resonances corresponding to this regime are shown in Fig.~\ref{fig:3d} for $P_p=1\,\mathrm{W}$. The experimental scans at this driving power confirm the shape of the nonlinear resonance and the existence of a bistable region for $\delta_2>0$ [see Fig.~\ref{fig:3e}].  

\begin{figure}
\customlabel{fig:4a}{4(a)}
\customlabel{fig:4b}{4(b)}
\customlabel{fig:4c}{4(c)}
\customlabel{fig:4d}{4(d)}
\customlabel{fig:4e}{4(e)}
\customlabel{fig:4f}{4(f)}
\customlabel{fig:4g}{4(g)}

\includegraphics[width=\columnwidth]{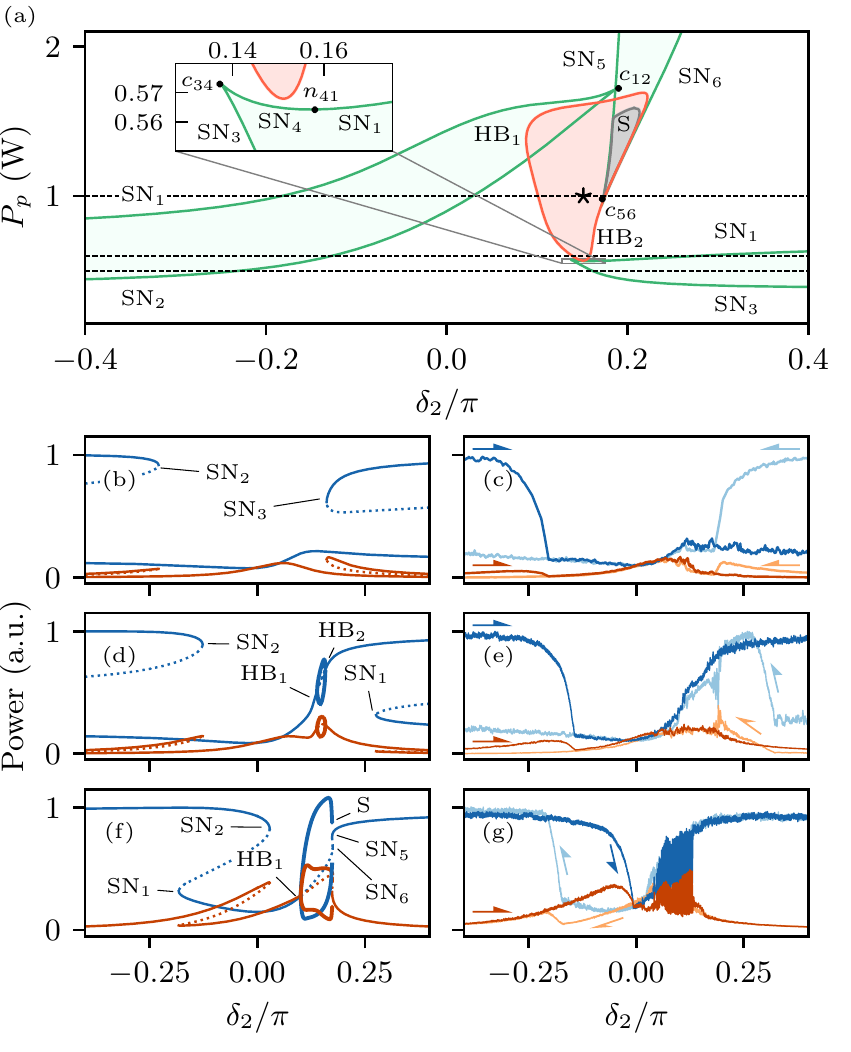}
\caption{Same as in Fig.~\ref{fig3_delta1-036}, but for $\delta_1=0.64$. (a) Phase diagram. The Hopf bifurcations $\mathrm{HB}_{1,2}$ (red lines) mark the boundaries of the self-pulsing region, and the Shil'nikov bifurcation S (gray line), the boundaries of the area where there is no self-pulsing. $n_{41}$, necking bifurcation; $c_{ij}$, cusp bifurcation. The star indicates the parameters of the experiment reported in Fig.~\ref{fig:5a}. (b, d, f) Normalized intracavity powers (thin lines) as a function of $\delta_2$ at a driving power of 0.5, 0.6 and $1\,\mathrm{W}$, respectively. The thick lines between the bifurcations $\mathrm{HB}_{1,2}$ show the maximum and minimum oscillation amplitudes. Note that in panel (f), $\mathrm{HB}_{2}$ (not shown for clarity) is located at $\mathrm{SN}_6$. The corresponding experimental scans are displayed in panels (c, e, g). See main text for further explanations.} 
\label{fig4-Bistab-SP}
\end{figure}
%
%
\begin{figure*}
\customlabel{fig:5a}{5(a)}
\customlabel{fig:5b}{5(b)}
\customlabel{fig:5c}{5(c)}
\customlabel{fig:5d}{5(d)}
\customlabel{fig:5e}{5(e)}
\centering
\includegraphics[width=\textwidth]{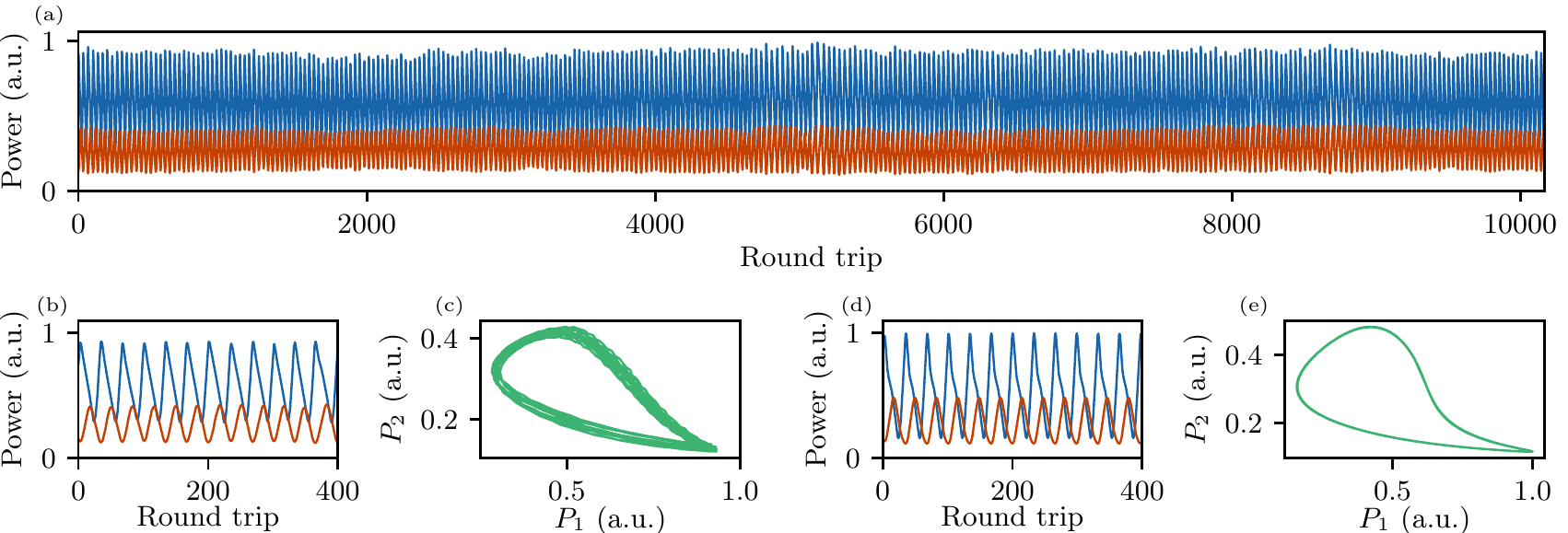}
\caption{(a) Normalized power in the driven (blue) and undriven (red) resonators, recorded over more that 10\,000 round trips ($12\,\mathrm{ms}$) and demonstrating periodic oscillations. The parameters are $P_p= 1\,\mathrm{W}$, $\delta_1=0.64$ and $\delta_2=0.47$. (b) Zoom over 400 round trips and (c) corresponding trajectory in the ($P_1,\,P_2$)-phase plane showing a limit cycle. (d) Sustained oscillations and (e) limit cycle obtained by numerical integration of Eq.~(\ref{MF2Ring}).  
}
\label{fig-exp-sp-amplitude}
\end{figure*}

The bifurcation analysis of the DDBH dimer model predicts the emergence of self-pulsing for larger detunings of the driven cavity. Such a situation is depicted in Fig.~\ref{fig:4a} for $\delta_1=0.64$, where the bistable regions as well as the self-pulsing area in the ($\delta_2,\,P_p$)-plane of the parameter space are plotted. 
At $P_p=500\,\mathrm{mW}$, the two saddle-node curves ($\mathrm{SN}_{2,3}$) are crossed when scanning $\delta_2$. We note that these two lines are actually connected as they stretch between adjacent resonances. At this power, the photonic dimer has two stable states for all $\delta_2$ values, except in the vicinity of the avoided mode crossing [see Fig.~\ref{fig:4b}]. However, since the upper state is completely disconnected from the lower one, the system cannot spontaneously jump to the high power state by changing the detuning of the undriven cavity. In order to observe the two states in the bistable region, we perturb $\delta_1$ to allow the system to switch to the higher state. It then falls back to the lower one while scanning $\delta_2$ in the forward or the backward direction. This is seen in Fig.~\ref{fig:4c}, where only the down-switching transition is reported for both scanning directions. The two $\delta_2$ values at which the switching takes place coincide well with the theoretical location of $\mathrm{SN}_2$ and $\mathrm{SN}_3$ in Fig.~\ref{fig:4b}. 

At a slightly higher driving power, two new saddle-node lines $\mathrm{SN}_1$ and $\mathrm{SN}_4$ appear as a necking bifurcation \cite{Prat2002} is crossed [see $n_{41}$ in the inset of Fig.~\ref{fig:4a}]. At this point, $\mathrm{SN}_3$ touches the lower state and a connection to the upper state emerges (not shown). It is beyond that power that self-pulsing occurs. The $\mathrm{SN}_4$ line however quickly disappears through a cusp bifurcation at $c_{34}$. Just above the power corresponding to $n_{41}$, $\mathrm{SN}_1$ and $\mathrm{SN}_2$ are located on either side of the newly formed connection between the lower and upper states. An example is given in Fig.~\ref{fig:4d} for $P_p=600\,\mathrm{mW}$. We show in the experimental scans [see Fig.~\ref{fig:4e}] that this leads to a hysteresis cycle that stretches between adjacent resonances. The system jumps to the lower state at $\mathrm{SN}_2$ in the forward direction, then goes back to the upper state thanks to the connection in the avoided crossing region. In the backward direction, it first switches to the upper state at $\mathrm{SN}_1$, then goes back down and stays on the lower state until the next resonance. Moreover, as seen in the experimental backward scan, the system spontaneously oscillates in a narrow range of $\delta_2$ which coincides with the self-pulsing region located between the two Hopf bifurcations $\mathrm{HB}_1$ and $\mathrm{HB}_2$. At a higher power, the self-pulsing region grows while $\mathrm{SN}_1$ moves away from the avoided crossing and, owing to the $2\pi$ periodicity, appears on the other side of the resonance. The stable and unstable homogenous solutions of Eq.~(\ref{MF2Ring}) are plotted for a $1\,\mathrm{W}$ driving power in Fig.~\ref{fig:4f}, as well as the boundaries of the oscillation amplitudes between the Hopf bifurcations. The measurements of the intracavity powers in the two resonators as $\delta_2$ is scanned back and forth are reported in Fig.~\ref{fig:4g}. They show a hysteresis cycle on the left and self-starting oscillations on the right, whose locations are in excellent agreement with the bifurcation analysis of Eq.~(\ref{MF2Ring}).

To prove that these oscillations correspond to self-pulsing, we next stabilize the second cavity around $\delta_2=0.47$, i.e. in the middle of the unstable region. A stable oscillation in both cavities is measured over $10\,000$ round trips [see Figs.~\ref{fig:5a} and~\ref{fig:5b}]. The trajectory in the ($P_1,\,P_2$)-phase plane reported in Fig.~\ref{fig:5c} shows a limit cycle behavior consistent with the numerical simulation of Eq.~(\ref{MF2Ring}) displayed in Fig.~\ref{fig:5e}. This confirms that the DDBH dimer model captures very well the nonlinear behavior of our system. 
We note that the deviations from the average orbit in the experiment are attributed to the limitations of the stabilization technique of the two cavity detunings.
The oscillation period is about 33.5 round trips, both in the experiment and the simulation. This value is larger than the one computed from the frequency splitting between the linear hybridized modes\,\cite{Zambon2020} (27.1 round trips). This difference can be explained by the weak coupling, since $\kappa \approx \sqrt{\theta_{12}}$ in our system.     

At a driving power close to 1\,W, the bifurcation diagram displayed in Fig.~\ref{fig:4a} shows that the system crosses a second cusp-bifurcation ($c_{56}$). It results in the emergence of two new continuous states separated by a pair of saddle-node bifurcations $\mathrm{SN}_{5,6}$. 
In-between these lines, the self-pulsing undergoes a Shil'nikov homoclinic bifurcation (S)\,\cite{glendinning_stability_1994,  Giraldo2020}. Here, the self-pulsing cycle is destroyed and an homoclinic orbit appears, leading to type II excitability of the photonic dimer\,\citep{izhikevich_neural_2012}. This is not reported because of a power limitation in our experiment. 
By further increasing the driving power, only $\mathrm{SN}_5$ and $\mathrm{SN}_6$ remain, similarly to the case reported in Figs.~\ref{fig:3d} and~\ref{fig:3e}.                

In summary, we have explored with fiber cavities the bifurcation structure of driven-dissipative Bose-Hubbard dimers. The ability to individually tune the two cavity detunings with respect to the driving field has been leveraged to study bifurcations in the ($\delta_2,\,P_p$)-plane of the parameter space for different $\delta_1$.  
We have shown that increasing the driven cavity detuning ($\delta_1$) dramatically modifies the hysteresis cycle while scanning the undriven cavity detuning ($\delta_2$).
Then, by driving the system out of equilibrium, we have observed the emergence of large amplitude oscillations leading to a well defined limit cycle. This self-pulsing behavior occurs in the vicinity of the avoided crossing, confirming the key role played by the coupling between the nonlinear sub-systems. 
All our results are very well explained by the canonical driven-dissipative Bose-Hubbard dimer model. Moreover, this model predicts the existence of a Shil'nikov bifurcation. On the one hand this bifurcation reduces the domain of existence of self-pulsing but, on the other hand, it enables excitability that may find applications in all-optical computing\,\cite{ Vandoorne2014, Shen2017,Feldmann2019}. The oscillations reported in this work involve a single resonance of the photonic dimer, or equivalently a single longitudinal mode. However, the synchronous oscillation of multiple modes may also occur. This has recently been predicted in pairs of coupled ring resonators with anomalous dispersion, giving rise to soliton hopping~\cite{Tikan2020}. Our results are therefore also relevant for the observation of the dynamics between solitons, in photonic dimers and in lattice of coupled resonators.       
  
The authors acknowledge fruitful discussions with Nicholas Cox and Emmanuel Gottlob. This work was supported by the Fonds de la Recherche Scientifique - FNRS under grant No PDR.T.0104.19 and the European Research Council (ERC) under the European Union’s Horizon 2020 research and innovation program (grant agreement No 757800). F.L., N.E. and P.P.-R. acknowledge the support of the Fonds de la Recherche Scientifique-FNRS).

\bibliography{PRL2}
    
\end{document}